\documentstyle[a4,epsbox]{article}

\def\l{\langle}
\def\r{\rangle}
\def\ps{| \psi_S \r}
\def\pp{| \psi_P \r}
\def\qi{| q_i \r}
\def\rj{| r_j \r}
\def\u{u_{ij}^{k \ell}}
\def\ud{u_{ij}^{k' \ell'}}
\def\c{c_{ij}(\vec a, \vec b)}
\def\hQ{{\hat Q}}
\def\hR{{\hat R}}
\def\hH{{\hat H}}
\def\dne{\delta n_{err}}

\begin{document}


\begin{figure}[t]
\vskip 13mm
\end{figure}

\begin{flushleft}
\large\bf 
First-kind measurements, non-demolition measurements, and conservation laws
\bigskip

\normalsize\rm
Akira Shimizu and Kazuko Fujita 
\smallskip

Institute of Physics, University of Tokyo, Komaba, Tokyo 153, Japan
\end{flushleft}
\medskip

\noindent{\bf Abstract}

A general discussion is given for first-kind 
(FK) and quantum non-demolition (QND) 
measurements. 
The general conditions for these measurements are derived, 
including the most general one (called the weak condition), 
an intermediate one, and the strongest one.  
The weak condition indicates that we can realize a FK or QND measuring 
apparatus of wide classes of observables 
by allowing the apparatus to have a finite response range. 
A recently-proposed 
QND photodetector using an electron interferometer is an example of such 
apparatus. 

\bigskip\noindent
{\bf 1.\ INTRODUCTION AND SUMMARY}
\medskip

\noindent
\qquad
When one measures an observable $\hQ$ of a quantum-mechanical system $S$,  
the measurement in general causes 
a quantum-mechanical back-action on $S$. 
As a result, he will not necessarily get the same value when 
he measures $\hQ$ again. 
A measurement $M_1$ is said to be of the {\it first kind} (FK) 
if a subsequent measurement $M_2$, 
made {\it immediately after} $M_1$, 
gives the same value of $\hQ$. 
On the other hand, $M_1$ is said to be 
of {\it quantum non-demolition} (QND)  
if $M_2$, made at {\it any later time} after $M_1$, 
gives the same value of $\hQ$ \cite{rev}. 
Any QND measurement is of the FK, 
but the inverse becomes true {\it only} when 
$\hQ$ is a constant-of-motion of $S$, 
i.e., when 
\begin{equation}
[\hQ, \hH_S] = 0, 
\label{QHS}\end{equation}
where $\hH_S$ is the hamiltonian of (isolated) $S$ \cite{rev}. 
It was argued \cite{LP} that 
a (quick) measurement can be of the FK 
{\it only} when 
\begin{equation}
[\hQ, \hH_I] = 0,   
\label{QHI}\end{equation}
where $\hH_I$ is the hamiltonian that describes the interaction 
between the measuring apparatus and $S$. 
Accordingly, the conditions for a QND measurement was 
often argued \cite{rev} to be the two conservation laws, Eqs.(1) and (2). 

If Eq.(2) were a {\it necessary} condition, 
then it would extremely restrict possible 
types of observables for FK or QND measurements \cite{LP}. 
However, several QND schemes which do {\it not} satisfy 
Eq.(2) have recently been proposed \cite{AS,eg}. 
The purpose of this work is 
to develop a general theory which 
can treat these examples and to clarify the underlying physics. 
It will be shown that Eq.(2) is a very strong, {\it sufficient} condition.  
We present general conditions including the most general one, which we call  
the weak condition, and an intermediate one.
The weak condition indicates that we can realize a FK or QND measuring 
apparatus of wide classes of observables 
by allowing the apparatus to have a finite response range. 
The QND photon-number counter proposed by the author 
\cite{AS} is an example of 
such apparatus. 
We also present additional conditions 
concerning the measurement error and the amount of obtained information. 

\bigskip\noindent
{\bf 2.\ MODEL OF MEASUREMENT PROCESSES}
\medskip

To calculate a state vector {\it after} a measurement, 
we shall use the projection ``postulate" of von Neumann, which  
in general terms is described as follows \cite{vN,Me}. 
When an ``ideal" measurement 
(which is sometimes called a ``moral" measurement \cite{De}) 
of $\hQ$ is performed on the 
system having the state vector $| \Psi \r$, 
the observed value will be an eigenvalue $q_n$ of $\hQ$, with 
the probability $P(q_n)$ given by the Born rule. 
In this case, the state vector {\it immediately} after the 
measurement is given by 
\begin{equation}
|\Psi_n' \r = |q_n \r \l q_n| \Psi \r/\sqrt{P(q_n)}
\quad \mbox{for an ideal measurement.}
\label{proj}\end{equation}
This is a precise mathematical description of the so-called 
``reduction of the wavepacket", and is 
called a ``postulate" because it 
is sometimes considered to 
lead to a conceptual (or philosophical) difficulty \cite{De}. 
Apart from such a conceptual problem, 
however, it is widely accepted that all experimental results must agree with 
theoretical 
predictions based on the Copenhagen interpretation, which assumes the 
wavepacket reduction \cite{De}. 
We therefore take Eq.(3), as standard textbooks do \cite{Me}, 
as one of the fundamental principles.  

Equation (\ref{proj}) holds only for an ``ideal" measurement which is 
defined as an error-less measurement of the first kind \cite{Me,De}. 
For this reason, it is sometimes argued that 
the standard quantum mechanics could not predict the state vector 
after a {\it non-ideal} measurement. 
However, this is false. 
Most -- probably any -- measurements can be treated 
by the coupled use of the principles 
[including Eq.(\ref{proj})] of quantum mechanics.  
A measurement is a series of many physical processes which 
subsequently take place in the measured system, measuring apparatus, 
and observer \cite{vN}. 
The point is that among 
the series of processes we can almost always
find an process which can be treated as 
an ideal measurement process. 
For example, suppose that the measured value is displayed on 
a digital display board of the apparatus. 
Then, reading the displayed number can be  
regarded, to a good approximation, as an ideal measurement of the number. 
So, we can apply Eq.(\ref{proj}) to this process. 
In this case, 
the measured system plus the apparatus must be treated as 
a coupled quantum system, and the processes occurring in this system 
should be analyzed by the Schr\"odinger equation. 
That is, between the display board and the observer's eyes is 
the boundary between the quantum system and the outer world 
which includes the observer.   
This allows us to calculate everything -- at least in principle. 
Although many candidates can usually be found for the boundary, 
the final results are invariant under different choices among the candidates, as shown by von Neumann \cite{vN}. 
The most practical choice is to take the boundary that leads to 
the smallest size of the quantum system. 

The above general consideration leads us to the model depicted in Fig.1. 
An observer measures an observable $\hQ$ of a quantum system $S$. 
We take the above-mentioned boundary somewhere in the measuring apparatus, 
and thereby 
decompose the apparatus into the ``probe" system $P$ and an ideal detector 
of an observable (read-out variable) $\hR$ of $P$. 
That is, $S$ plus $P$ constitute a coupled quantum system, 
with an interaction hamiltonian $\hH_I$, 
and the rest is the outer world. 
The observer will look at the detector of $\hR$ (probably with the help of 
some other apparatus), and he will estimate the value of $\hQ$ from 
the observed value of $\hR$, whereupon a measurement error usually enters
(see section 5). 

\begin{figure}[h]
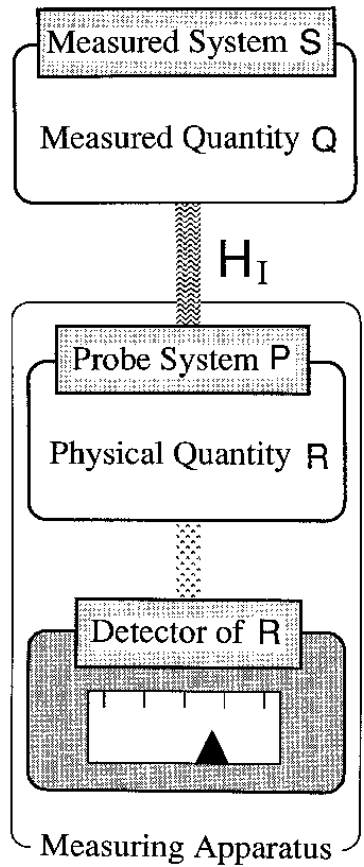

\begin{center}
\epsfile{file=Fig1,scale=0.4}
\end{center}
\caption{
Model of measurement
}
\end{figure}

\bigskip\noindent
{\bf 3.\ EVOLUTION OF THE COUPLED QUANTUM SYSTEM}
\medskip

As mentioned in section 1, any QND measurement is of the FK, 
and the inverse can be easily judged from Eq.(\ref{QHS}), 
which is related with the unitary evolutions 
{\it before} and {\it after} the measurement. 
We therefore need to focus only on the evolution {\it during} 
the measurement, and 
compare the state vectors {\it just before} and {\it just after} 
the measurement. (Henceforth the word ``just" will be omitted.)
This allows us to treat FK and QND on an equal footing. 

We assume that the premeasurement state vector $| \Psi \r$ 
of the coupled quantum system $S+P$ is 
the product of the state vectors of the subsystems: 
\begin{equation}
| \Psi \r = \ps \pp. 
\label{pre}\end{equation}
Let us expand $\ps$ and $\pp$ in terms of eigenfunctions 
of $\hQ$ and $\hR$, respectively: 
\begin{equation}
\ps = \sum_i a_i \qi, 
\quad 
\hat Q ~\qi = q_i \qi,
\label{ps}\end{equation}
\begin{equation}
\pp = \sum_j b_j \rj, 
\quad 
\hat R ~\rj = r_j \rj.  
\label{pp}\end{equation}
The states $\ps$ and $\pp$ are thereby 
expressed by the vectors $\vec a$ and $\vec b$, respectively. 
When interaction $\hH_I$ is switched on, 
a state $|q_k \r |r_\ell \r$  
undergoes a unitary evolution, which we write as 
\begin{equation}
|q_k \r |r_\ell \r 
\quad \rightarrow \quad
\sum_{i,j} \u \qi \rj, 
\end{equation}
where the unitary matrix $\{ \u \}$ is a function of $\hH_I$. 
From the superposition principle, 
$| \Psi \r$ evolves like 
\begin{equation}
| \Psi \r 
\quad \rightarrow \quad
| \Psi' \r = 
\sum_{i,j} \c \qi \rj, 
\quad 
\c \equiv \sum_{k, \ell} a_k b_\ell \u.  
\end{equation}
When $\hR$ is measured by the detector, 
the probability for getting the value $R=r_j$ is given by the Born rule; 
\begin{equation}
P(r_j) = \l \Psi' | r_j \r \l r_j | \Psi' \r = \sum_i |\c|^2, 
\end{equation}
and, according to Eq.(\ref{proj}), the state vector 
undergoes the non-unitary evolution; 
\begin{equation}
| \Psi' \r 
\quad \rightarrow \quad
| \Psi_j'' \r = 
\rj \l r_j | \Psi' \r / \sqrt{P(r_j)}
\quad \mbox{when the observed $R$ is $r_j$.}
\end{equation}
It is convenient to consider the density operator 
of the mixed ensemble of various post-measurement states 
$| \Psi_j'' \r$ corresponding to 
different observed values of $R$. 
This operator is given by 
\begin{equation}
\hat \rho'' = 
\sum_j P(r_j) | \Psi_j'' \r \l \Psi_j'' | = 
\sum_j \l r_j | \Psi' \r \l \Psi' \rj \cdot \rj \l r_j |. 
\label{post}\end{equation}

\bigskip\noindent
{\bf 4. GENERAL CONDITIONS FOR FK OR QND MEASUREMENT}
\medskip

From the definition of FK and QND measurements, 
it can be shown that the statistical distribution of $Q$ is invariant 
under these measurements. 
This invariance is very characteristic of these specific measurements, 
and thus the literature often defined QND measurements 
by this invariance. 
[For completeness, we must also require additional conditions 
on the measurement error and the amount of obtained information, which 
will be discussed in the next section.] 
The invariance can be expressed, 
with the help of Eqs.(\ref{ps}) and (\ref{post}), as  
\begin{equation}
\sum_{j} 
\left| \c \right|^2 = |a_i|^2 
\quad \mbox{: \ weak condition}.
\label{weak}\end{equation}
This is the most general condition for FK or QND measurement, 
which we call the weak condition. 
Both sides of this equation contain $\vec a$, the state of $S$. 
It is therefore possible that some measurement satisfies 
the weak condition {\it only for} particular states of $S$. 
That is, 
the weak condition includes such a case 
that a measurement becomes of the FK or QND only for a particular 
set of {\it measured} states. 
This corresponds to a limitation of response range of the measuring apparatus. 
Note that {\it any} 
existing apparatus do have finite response ranges, and thus 
the above possible limitation is quite realistic. 
Actually, as will be discussed below, only by accepting such a limitation 
can we realize FK or QND measurement for wide classes of observables. 

Before discussing this point, let us 
derive a condition which does not contain $\vec a$. 
We call it the moderate condition because 
it is a stronger condition than Eq.(\ref{weak}). 
It is given by 
\begin{equation}
\sum_{j} 
\left( \sum_{\ell} \u b_\ell \right)^*
\left( \sum_{\ell'} \ud b_{\ell'} \right)
= \delta_{ki} \delta_{k'i}
\quad \mbox{: \ moderate condition.}
\label{moderate}\end{equation}
The left-hand side of this equation contains $\vec b$, the state of $P$. 
It is therefore possible that some measurement satisfies 
the moderate condition {\it only for} particular states of $P$. 
That is, 
the moderate condition includes such a case 
that a measurement becomes of the FK or QND only for a particular 
set of {\it probe} states. 
The probe state can be prepared at will, at least in principle. 
It is a matter of design of the apparatus. 
[Recall that $P$ is a part of the apparatus.]  
The limitation on the probe states is therefore quite acceptable. 

We can also derive an even stronger condition, 
which we call the strong condition and is given by 
\begin{equation}
\u ~\propto~ \delta_{ki}
\quad \mbox{: \ strong condition.}
\label{strong}\end{equation}
It does not contain either $\vec a$ or $\vec b$. 
Hence, a measurement which satisfies the strong condition 
is of the FK or of QND for {\it any} measured states and probe states. 
We can easily show that Eq.(\ref{strong}) is equivalent 
to Eq.(\ref{QHI}) if Eq.(\ref{QHS}) is satisfied. 
That is, in our terms, 
the QND condition given in Ref.\cite{rev} 
is the strong condition. 
Any measurement which satisfies the strong (moderate) condition 
satisfies the moderate (weak) condition, but the inverse is not necessarily 
true. 

We now discuss implications of these conditions. We first 
note that a microscopic form, which describes 
elementary processes, should be used for $\hH_I$. 
Previous work frequently used effective forms, 
which approximately describe 
effective interactions resulting from many elementary processes. 
However, most work did not investigate whether such an effective 
interaction could correctly 
describe quantum-mechanical backactions. 
It should be emphasized that 
if one studies the problem correctly, that is, if he 
employs a microscopic $\hH_I$, then he will find that 
most observables of interest, such as a photon number, 
can {\it never} satisfy Eq.(\ref{QHI}) or Eq.(\ref{strong}) \cite{LP}. 
That is, for most observables of interest, 
FK or QND measurements are possible only in the moderate or weak sense. 
Therefore, it is extremely important for any QND proposals to clarify, 
using a microscopic interaction Hamiltonian,  
the limitations on the allowable measured states and/or the probe state. 
Unfortunately, however, such analysis is lacking in most work. 

Reference \cite{AS} did perform 
complete theoretical analysis on a QND photodetector. 
From the present viewpoint, what was found in 
the analysis is that the detector satisfies {\it neither} 
the strong condition {\it nor} the moderate condition. 
The detector satisfies only the weak condition: both the 
allowable measured states and the probe state are limited. 
That is, the measured photon states must be in a pulse (wavepacket) 
form with a smooth envelope whose width is long enough (typically 
10 ps) so that Eq.(3) of Ref.\cite{AS} is satisfied. 
For shorter pulses, the detector no longer works as a QND detector. 
The probe of the detector is electrons. 
The electrons must be confined in one-dimensional 
quantum wires. Otherwise, they would scatter or 
absorb photons, and the detector would not work as a QND detector. 

Note also that Eq.(\ref{weak}) is much more general than the 
following condition suggested by Vaidman \cite{Va}; 
\begin{equation}
[\hQ, \hH_I] \ps = 0.  
\end{equation}
In fact, the QND photodetector of Ref.\cite{AS} does not satisfy 
this equation. 
The points are that (i) in general $\u$ cannot be 
reduced to such a simple form, and (ii) Eq.(\ref{weak}) contains 
the information on the probe state, whereas the above equation not. 

\bigskip\noindent
{\bf 5.\ ESTIMATION OF Q FROM OBSERVED R}
\medskip

We now discuss additional conditions for FK or QND measurements, concerning 
the measurement error and the amount of 
information obtained by the measurement. 
Let $r$ be the observed value of $R$. 
Since the observer knows the mechanism of his apparatus, he can  
estimate the value of $Q$ from $r$.
The estimation may be expressed in a functional form as 
$q_{est} = f(r)$, where $q_{est}$ is the estimated value of $Q$. 
Using this function, we define the operator for the estimated value by 
$
\hat Q_{est} \equiv f(\hat R).
$
In order for the estimation to be a good one, 
it is required that 
$
\l \hat Q_{est} \r = \l \hat Q \r, 
$
where $\l \cdots \r$ denotes the expectation value in the 
post-measurement states, Eq.(\ref{post}). 
[Note that 
the expectation value of $Q^n$ ($n = 1, 2, \cdots$) for the pre-measurement 
state is equal to that for the post-measurement state, because, 
as mentioned in the previous section, 
the distribution of $Q$ is invariant under FK or QND measurements.]
This requirement can be rewritten as
\begin{equation}
\sum_{i,j} 
\left| c_{ij}(\vec a, \vec b) \right|^2 
[q_i - f(r_j)]=0. 
\end{equation}
It is also required that the measurement error must be small enough:  
$
\delta Q_{err}^2 \equiv 
\l (\hat Q_{est} - \hat Q )^2 \r 
\ \leq \ \epsilon^2, 
$
where $\epsilon$ is the maximum allowable error. 
This requirement can be rewritten as 
\begin{equation}
\sum_{i,j} 
\left| c_{ij}(\vec a, \vec b) \right|^2 
[q_i - f(r_j)]^2 
\ \leq \ \epsilon^2.
\end{equation}
For example, the measurement error of the QND photodetector of Ref.\cite{AS}
is inversely proportional to the number of probe electrons, and 
thus can be made arbitrarily small -- at least in principle.  

We also require that the amount of information $I$ obtained by the measurement 
must be large enough: 
$
I \geq I_{min}, 
$
where $I_{min}$ is the minimum allowable amount of information. 
Although this last requirement was frequently disregarded, 
it is an essential requirement for a physical process to be called 
a measurement. For example, if the observer knows the premeasurement state 
vector beforehand, he can evaluate expectation values of any observables 
without performing actual measurement, i.e., without causing any change 
in the measured state vector. 
Can it be called a QND measurement? 
Since he knows the state vector beforehand, 
the amount of information he can get through his ``measurement" is 
$
I = - 1 \ln 1 = 0. 
$
Hence, his ``measurement" is by no means a measurement. 
By contrast, 
$
I 
\simeq \ln (n_{max}/\dne)
$
for a QND photon counter with 
the response range $n \leq n_{max}$ and the measurement error $\dne$.
For the QND photodetector of Ref.\cite{AS}, for example, 
typical values are $\dne \sim 10^2$ and $n_{max} \sim 10^6$, so that 
$I \simeq 12$, which is comparable with usual non-QND photodetectors. 
We can make $\dne$ smaller by confining photons in a cavity \cite{AS}, 
which leads to even larger $I$. 

\vspace{-3mm}
\small

\end{document}